\documentclass[copyright,creativecommons]{eptcs}
\providecommand{\event}{FAVO 2009} 
\providecommand{\volume}{??}
\providecommand{\anno}{20??}
\providecommand{\firstpage}{1}
\usepackage{graphicx}
\usepackage{fpcsp,zed-csp}
\providecommand{\eid}{??}
\usepackage{breakurl}        

\title{Formal Modelling of a Usable Identity Management Solution for Virtual Organisations}
\author{Ali N. Haidar, P. V. Coveney
\institute{Centre for Computational Science\\ University College London\\ London, UK}
\email{\{Ali.Haidar, P.V.Coveney\}@ucl.ac.uk}
\and
Ali E. Abdallah
\institute{E-Security Center\\ Department of Informatics\\
London South Bank University, UK}
\email{A.abdallah@lsbu.ac.uk}
\and
P. Y. A. Ryan
\institute{University of Luxembourg\\
Luxembourg}
\email{peter.ryan@uni.lu}
\and
B. Beckles
\institute{University of Cambridge Computing Service\\
Cambridge, UK}
\email{\quad mbb10@cam.ac.uk}
\and
J. M. Brooke, M . A. S. Jones
\institute{University of Manchester\\
Manchester, UK}
\email{\{mike.jones,john.brook\}@manchester.ac.uk}
}
\def\titlerunning{Formal Modelling of a Usable Identity Management Solution for Virtual Organisations}
\def\authorrunning{Ali N. Haidar et al.}
\begin{document}
\maketitle

\begin{abstract}
This paper attempts to accurately model security requirements for computational grid environments with particular focus on authentication. We introduce the Audited Credential Delegation (ACD) architecture as a solution to some of the virtual organisations (VO) identity management usability problems. The approach uses two complementary models: one is state based, described in Z notation, and the other is event-based, expressed in the Process Algebra of Hoare's Communicating Sequential Processes (CSP). The former will be used to capture the state of the VO and to model ``back-end"
operations on it whereas the latter will be used to model behavior, and in particular, ``front-end" interactions and communications.
The modelling helps to clearly and precisely understand functional and security requirements and provide a basis for verifying that the system meets its intended requirements.
\end{abstract}

\section{Introduction}
The mission of a Virtual Organisation (VO) is to offer a simplified end user access to and use of high performance computing resources shared across a number of different institutions with different administrative security domains. A typical example of a VO is the computational grid, which aims to provide control over distributed resources consisting of enormous computational power (parallel processing machines), data storage (hard disks, memory) and visualisation on high speed networks. Examples of currently operating grids include: the UK National Grid Service (NGS) \cite{NGS} and US TeraGrid \cite{TerraGrid}. The sharing of these resources is intended to support academic research and industrial development. A computational grid environment may consist of a mixture of several kinds of organisations including academic, governmental, industrial and commercial institutions (will be referred to as ``Sites" in this paper).

\noindent One major problem faced by end-users and administrators of VOs is to do with the usability of the security mechanisms usually deployed in these environments \cite{GridProject2006,JIAS2009}, in particular identity management solutions. Many of the existing computational grid environments use Public Key Infrastructure (PKI) and X.509 digital certificates \cite{Cho02} as a corner stone for their security architectures. However, security solutions based on PKI have to be usable to be effective. Some of the common grid identity management encountered include: adding and removing users, acquiring and using digital certificates. End-users, such as scientists who are not security experts, are concerned with the results of the computations they perform on the grid. Many of the existing grid middleware, such as Globus and Unicore \cite{GridProject2006}, require that the end-user creates a short lived certificate, known as proxy certificate, prior to running application on the grid. In addition, many users engage in practices which weakens the security of the grid environment, such as the sharing of the private key of a single personal certificate because acquiring certificates can be a lengthy process \cite{JIAS2009}.

We introduce the Audited Credential Delegation (ACD) architecture as a solution to some of the above problems and we present a formal model of this architecture. The proposed solution carefully hides the use of digital certificates from end-users and enables them to acquire credentials from their local site administrators. It also enables the latter to to create/remove user accounts in a more efficient way. A combination of state-based model,  described in Z notation \cite{jim96}, and event-based model, expressed in the Process Algebra of Hoare's Communicating Sequential Processes (CSP) \cite{hoa85} is used to model the architecture. Z is widely used in industry for modelling complex and large systems. It  will be used to capture the state of the VO components and to model ``back-end" operations on it whereas the latter will be used to model behavior, and in particular, ``front-end" interactions and communications. Both notations have clear and precise semantics. The Z descriptions in this work have been type checked with ZTC tool. The modelling helps to clearly and precisely understand functional and security requirements and provide a basis for verifying that the system meets its intended requirements. There are several formal frameworks that combine state-based and event-based approaches and can also be used to  have a clear and concise model of this architecture, such as Circus in \cite{CPA04Woodcock} and CSP$\parallel$B in \cite{Shneider2005a}. Circus also combines Z and CSP. In this work we are interested in modelling the VO architecture and the nature of Z as a pure specification language \cite{jim96}, with a purer mathematical notation is therefore more appropriate than VDM and B because they are more akin to conventional programming languages, and hence why refinement into code is easier with these languages.

\noindent The remainder of this paper is organized as follows. Section 2 gives a brief overview of the proposed  VO architecture. Section 3 and 4 present formal state-based models of the authentication components followed by a CSP description of their pattern of interactions. Section 5 presents our conclusion.

\section{Overview of Proposed Architecture}
The physical infrastructure of this VO involves a separate administration site and a dedicate gateway service, which is aimed at hiding digital certificates from end-user environment.

\begin{figure}[!htbp]
\centering
\includegraphics[width=3.6in]{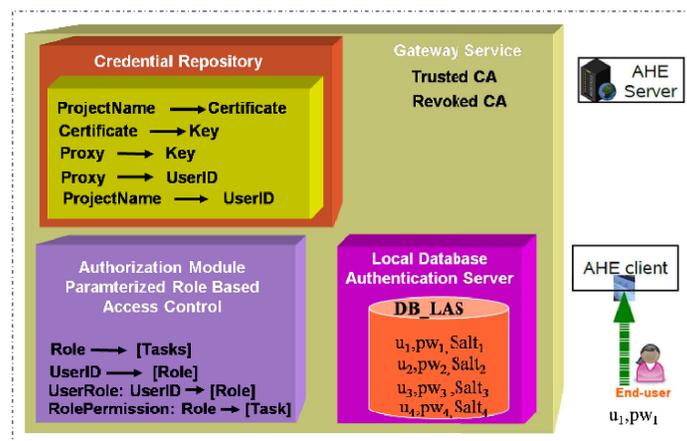}
\caption{Audited Credential Delegation Architecture} \label{Architecture}
\end{figure}

\noindent The gateway is responsible for identity management and consists of the following components: (1) Credential repository that stores certificates and their corresponding private keys in order to communicate with the computational grid. It also maintains a list of active proxy certificates, their corresponding privates key and an association between users and proxies. The main role of this component is to enable local site users to authenticate to external sites in the grid. (2) Local Authentication Service (LAS) component that enables authenticating local site users within their organization using a locally provided usable authentication mechanism rather than digital certificates. The LAS can support several types of authentication mechanisms that scientists are used to such as Kerberos \cite{Gol99}, Shibboleth \cite{GridSecureAccess06} or a local password database maintained at the gateway. End-users interact only with this component of the gateway. (3) An authorization component that controls requests issued from local site users to Grid resources. In this paper the focus is on the local and external authentication components. The gateway will be integrated with the Application Hosting Environment (AHE) \cite{AHE2007}, which allows scientists to run application codes on grid resources; manage the transfer of files to and from the grid resources; and allow the user to monitor the status of application instances that are run on the grid resources. This way it becomes possible to identify legitimate users and to ensure that only those legitimate users are allowed access to grid resources according to the policy defined on the grid projet. In this model, a scientist can login locally using a username/password pair for the whole session and run applications on the grid via the AHE server in a controlled manner. We are hoping to be able to deploy this solution within AHE for use on TeraGrid, NGS and DEISA within the next 12 months.

\section{VO Internal Authentication}
The aim of this component is to enable end-user to authenticate locally using a username-password mechanism. To be authenticated, a user must show
knowledge of a valid username/password pair that matches an entry in an authentication table, which can be a database or a password file.  In this work, a simple database password is considered for simplicity. In future work, Shibboleth and Kerberos can also be used in the architecture. One of the techniques used for implementing this approach is to store the hash of a salted password rather than the password itself in clear text. This way it is possible to know that the user knows the correct password without ever having to store the original password on the authentication server.

\begin{figure}[!htbp]
\centering
\includegraphics[width=4.0in]{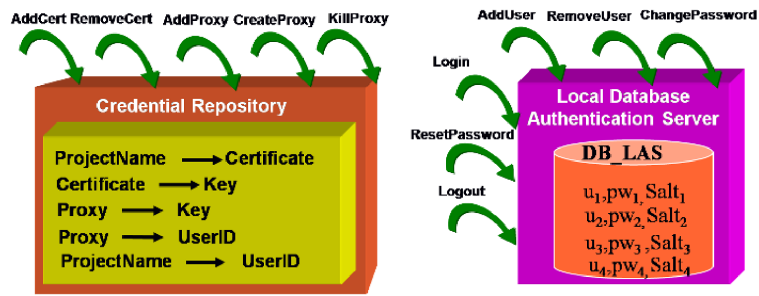}
\caption{Local Database and Credential Repository that store digital Certificates, proxies and users' credentials} \label{CertificateRespository}
\end{figure}

\subsection{State-Based Model of the Local Authentication Component}
Let \emph{UserID} and \emph{Data} be abstract types for denoting the set of
all usernames, passwords and encrypted passwords.
\begin{zed}
     [UserID, Data]
\end{zed}

\noindent \textbf{State}:  The state of the local database authentication server comprises: a set of
registered users; a partial function $pwdDB$ that associates each
$userID$ with one encrypted password; partial function that associates each user with a salt; and a partial function
$encrypt$ that is used to encrypt/hash clear text passwords. The
invariant ensures that every registered user must have a password and that every user has an associated salt.
The model can be described in \emph{Z} as follows:

\begin{schema}{DB\_LAS}
registered\_users : \power UserID \t1 pwdDB : UserID \pfun Data \\
salting: UserID \pfun Data \t1 encrypt: Data \pfun Data
\where
registered\_users = \dom pwdDB \land \dom pwdDB = \dom salting
\end{schema}

\noindent \textbf{Authentication Component Operations}: The set of operations considered on this component are shown in Figure \ref{CertificateRespository}. We only present the following operations on the $DB\_LAS$: \emph{Login}, \emph{ChangePassword} and \emph{AddCredential}because of space limitation. For more details about this model the reader is referred to \cite{HaidarA08}.

\noindent The operation $Login$ takes a username and a password as inputs and checks whether the pair matches an entry in
the database. The operation is described in the following \emph{Z} schema:

\begin{schema}{Login0}
     \Xi DB\_LAS \\
     username? : UserID \t1     pwd? : Data
   \where
   encrypt(salting(username?)+pwd?) = pwdDB(username?)
\end{schema}

\noindent The operation $ChangePassword$ replaces the old
password for the specified $username$ with a new password after
checking that the username and the old password supplied by a user matches an entry
in the database:

\begin{schema}{ChangePassword0}
    \Delta DB\_LAS \\
     username? : UserID \t1    pwd? : Data \t1  newpwd? : Data \t1  newsalt? Data
 \where
    encrypt(salting(username?)+pwd?) = pwdDB(username?) \land \\
    pwdDB'= pwdDB \oplus \{username? \mapsto encrypt(newsalt?+newpwd?)\} \\
    salting' = salting \oplus \{username? \mapsto newsalt? \}
 \end{schema}

\noindent The $AddCredential$ operation allows adding a new user to the database.

 \begin{schema}{AddCredential0}
    \Delta DB\_LAS \\
    username? : UserID \t1  pwd? : Data \t1    salt?: Data
\where
     pwdDB' = pwdDB \oplus \{username? \mapsto encrypt(salt?+pwd?)\} \\
     salting' = salting \oplus \{username? \mapsto salt? \}
\end{schema}

\noindent Let $Report$ be a data type, the values of which are
messages indicating whether an operation has been successful or has
failed.
\begin{syntax}
    Report & ::= & Success | Failure
\end{syntax}

\noindent \textbf{Precondition of each Operation}: The precondition of the operations $Login$ and $ChangePassword$ is that the username and password pair match an entry in the database. The precondition of the operation $AddCredential$ is that the chosen username must not be already in use. The precondition for each operation can be defined in $Z$ as follows:
\begin{zed}
 \pre Login \defs (username?,encrypt(salting(username?)+pwd?)) \in  pwdDB \\
\pre ChangePassword0 \defs (username?,encrypt(salting(username?)+oldpwd?)) \in  pwdDB \\
\pre AddCredential0 \defs username? \notin \dom pwdDB \\
\end{zed}

\noindent \textbf{Totalizing}: The totalising technique is used to handle errors resulting from not meeting the above preconditions.
An error may arise because the $username$ doesn't exist,
\begin{schema}{UserIDNotInUse}
\Xi DB\_LAS \\
username? : UserID \t1 athrep!: Report
\where username? \notin \dom pwdDB \land athrep! = Failure
\end{schema}

\noindent or the combination of username and password is wrong:
\begin{schema}{InvalidCredential}
 \Xi DB\_LAS \\
username?: UserID \t1 pwd?: Data \\
authenticationdecision!: Report
\where \lnot (encrypt(pwd?) = pwdDB(username?)) \land authenticationdecision! = Failure
\end{schema}

\noindent A successful operation will result in the same report:
\begin{schema}{Op\_Success}
authenticationdecision! : Report \where authenticationdecision! = Success
\end{schema}

\noindent The Authentication component's operations will then be
modeled as follows:
\begin{zed}
    Login \defs (Login0 \land Op\_Success) \lor InvalidCredential \also
    ChangePassword \defs (ChangePassword0 \land Op\_Success) \lor UserIDNotInUse  \lor InvalidCredential \also
    AddCredential  \defs (AddCredential0 \land Op\_Success) \lor UserIDInUse  \also
\end{zed}

\noindent \textbf{Initialization}: The initial state of the authentication server
component is described as follows:
\begin{schema}{DB\_LASInit}
AuthenticationCredential'
 \where
   registered\_users' = \{ali, mark, john\} \\
   pwdDB' = \{(ali,6f8cac5b994687f7a05619c3324fbc5e),\\
   (mark,8d137ac4eec0df89f089540ac19ac99c),(john,a4375b7cc7511652d0029cbffff4269) \}
\end{schema}

\noindent In this initialization, the hashes of the passwords are generated using MD5 hash \cite{MOV97}. The username/password pairs memorized by the users are: $(ali,pwdx),(mark,mrk3000),(john,wnd1980)$

\subsection{Modelling the User}
The model of a user focuses primarily on the security knowledge
that the user must possess and maintain for the purpose of
authentication. The abstract state of a user, $User$, comprises
three components: (1) $u\_names$, set of usernames held by the user (also known as principal); and (2)
$u\_password$, a function that associates each principal with a
plain password. The state of a user can be formulated in $Z$ as
follows:

\begin{schema}{User}
u\_names: \power UserID \t1 u\_password: UserID \pfun Data
\where \dom u\_password =
u\_names
\end{schema}
\noindent The invariant states that each username has exactly one password.

\subsection{Event-Based Model of the Local Authentication Component}
We derive the CSP interface of the authentication server from the \emph{Z} specification. This description has been structured as follows:
\begin{itemize}
  \item State =  DB\_LAS.
  \item Operations = \{Login , ChangePassword, ResetPassword, AddCredential,RemoveCredential,Logout\}.
  \item Preconditions of each operation. For instance, for the login operation:\\
   \t1 $Login$: $(username?,encrypt(salting(username?)+pwd?)) \in pwdDB$
  \item Initial WS state =  DB\_LASInit, totalizing the operations and adding reports to handle incorrect inputs.
\end{itemize}

\noindent  The interface of the authentication server is:
\begin{eqnarray*}
&& \alpha DB = \{Login, LoginRequest,LoginResponse,  ChangePassword, ChangePasswordRequest,\\
&&  ChangePasswordResponse, ResetPassword, ResetPasswordRequest,ResetPasswordResponse, \\
&&  AddCredential,AddCredentialRequest,AddCredentialResponse, RemoveCredential,\\
&&  RemoveCredentialRequest,RemoveCredentialResponse, Logout, LogoutRequest, LogoutResponse \}
\end{eqnarray*}

\noindent The behaviour of the authentication server is modelled by the CSP process $DB$ shown below. The description doesn't consider the pattern of interactions corresponding to the  operations $ResetPassword$, $RemoveCredential$ and $AddCredential$, because this depends on the role of the authenticated user. For example, only an authenticated user holding an administrator role can add/remove other users and reset passwords.

\begin{eqnarray*}
&& DB(State)  =  LOGIN(state)  \\
&& LOGIN (state)  =  Login \then LoginRequest?(username?,pwd?) \then \\
&& \t0 (\thenifelse{LoginResponse!(authenticationdecision!! \bindsto Success)\\
&& \t1\then AUTH(u)} {\pre Login(username?,pwd?)} {} \\
&& ~ LoginResponse!(authenticationdecision! \bindsto Failure) \then DB((state))) \\
&& AUTH(u)  = CHGPWD (state) \extchoice LOGOUT (state) \\
&& CHGPWD (state) = ChangePassword \then \\
&& ChangePasswordRequest?(username?,oldpwd?,newpwd?) \then\\
&& \t0  (\thenifelse{ChangePasswordResponse!(athrep! \bindsto Success) \then \\
&& \t1 AS(state')} {\pre ChangePassword(username?,oldpwd?,newpwd?)} {} \\
&&   ChangePasswordResponse!(athrep! \bindsto Failure) \then AUTH(u)) \\
&& LOGOUT (state) =  Logout \then LogoutRequest?(username?) \then \\
&& (\thenifelse{LogoutResponse!(athrep! \bindsto Failure) \then AS(state')} {\pre Logout(username?)} {} \\
&&  \t0 LogoutResponse!(athrep! \bindsto Failure) \then DB(state))
\end{eqnarray*}

\begin{figure}[!htbp]
\centering
\includegraphics[width=3.5in, height =0.8in]{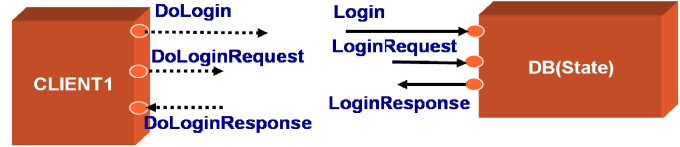}
\caption{Client and Database Server models} \label{Client}
\end{figure}

\noindent So for example, a user with a valid username and password,
say $ali$ and $pwdx$ respectively, can be authenticated by the
server by issuing the following sequence of interactions:

\begin{eqnarray*}
&& CLIENT1  =   DoLogin \then DoLoginRequest!(username? \bindsto ali, pwd? \bindsto 6f....fbc5e)   \\
&& \then DoLoginResponse?authenticationdecision! \then SKIP
\end{eqnarray*}
\noindent where $encrypt(pwdx) = 6f8cac5b994687f7a05619c3324fbc5e$.\\

\noindent The CSP operator $ [+Op+] $ models the interaction between two processes in which the handshake is on the operation $Op$. Both processes synchronize on the channel $Op$. The input values flows from the requestor to the provider and the output values flows in the opposite direction. For instance, the result of $CLIENT1$ sequence of interactions with the authentication server is calculated using a parallel composition of
$CLIENT1$ and $AS$ processes as follows:

\begin{eqnarray*}
&& DB(state) [+Login+]  CLIENT1 = Login \then \\
&& LoginRequest!(username? \bindsto ali, pwd? \bindsto 6f8cac5b994687f7a05619c3324fbc5e)   \\
&& LoginResponse!(authenticationdecision! \bindsto Success) \then DB(state')
\end{eqnarray*}

\begin{figure}[!htbp]
\centering
\includegraphics[width=3.5in]{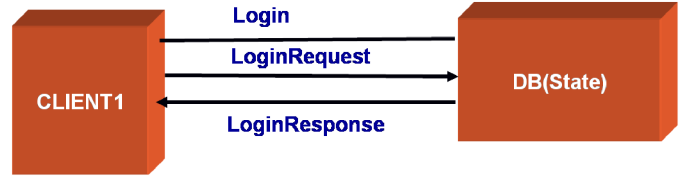}
\caption{Client and Database handshake on Login Operation} \label{ClientServer}
\end{figure}

\section{VO External Authentication}
The credential repository is used to store digital certificates or proxy certificates for named grid projects and resources and their corresponding private keys to enable communication with the grid. These certificates will be shared by a group of users and will be used to create proxy certificates.

\subsection{State-Based Model of the External Authentication Component}
The model assumes the existence of the following
types:

\begin{zed}
     [UserID, Subject, Data, Key, serialNb, CipherAlgName,CertAuthorityName]
\end{zed}

\noindent The state of the certificate repository comprises: a set of
certificates; a set of project and resources names; a partial function $key\_association$ that associates each
$Certificate$ with it's corresponding private key; a partial function
$cert\_association$ that is used to associate each project with a digital certificate; a list of issued proxies certificates created using the digital certificates, the proxies secret keys, association between each proxy and the user who generated it.\\

\begin{schema}{CredentialRepository}
    certificates : \power Certificate \t1  proxyCertificates: \power Certificate \\
    projectsNames: \power Name \t1 resourcesNames: \power Name \\
    key\_association : SerialNb \rel Key \t1 cert\_association:  Name \pfun SerialNb \\
    issuedproxies : SerialNb \pfun SerialNb \t1   proxyIssuer: SerialNb \pfun SerialNb\\
    proxySecretKey: SerialNb \pfun Key \t1  userProxy: SerialNb \pfun UserID
 \where
 \forall c: Certificate @ c \in certificates \land c.serial \in  \dom key\_association \land\\
   \ran cert\_association \subseteq \dom key\_association  \land\\
   \dom cert\_association \subseteq  projectsNames \land \\
    \dom cert\_association \subseteq resourcessNames \land \\
  \forall c: Certificate @ c \in proxyCertificates \land c.serial \in  \dom proxySecretKey \land \\
   \dom proxySecretKey = \dom proxyIssuer = \dom userProxy\\
\end{schema}

\noindent Where $proxyCertificates$ is the set of all active proxies; $issuedproxies$, a function that relates a serial number to a proxy certificate; $proxyIssuer$, a function that relates the proxy certificate with its issuer (signer) identified by a public certificate;  $userProxy$, a function that associates a user in a site with the proxy certificate in a unique way; $proxySecretKey$, a function that associates each proxy with its corresponding private key. More details on modelling PKI component in Z and CSP  are presented in \cite{HaidarA08C}.The same approach as in the previous section can be applied to model operations. For instance, the administrative operation on the
$CredentialRepository$, $AddCertificate$, takes a certificate, its corresponding private key, and the project with which it can be used as inputs. The precondition for this operation to succeed states that the $cert?$ should not already be in the list of
$certificates$. The operation is captured in Z as follows:

\begin{schema}{AddCertificate0}
    \Delta CredentialRepository \\
    cert? : Certificate \t1   secretkey? : Key \t1  project? : Name \\
    response!: Report
\where
    cert!.publicKey \inrel{validPKIKeyPair} secretKey? \land \\
    cert? \notin  certificates \land certificates' = certificates \cup \{cert?\} \land \\
     key\_association' = key\_association \cup \{ (cert?.serial, secretkey?) \} \land\\
    cert\_association' = cert\_association \oplus \{ project? \mapsto cert?.serial \}
\end{schema}

\noindent The operation $RemoveCertificate$ takes a certificate as an argument
and removes it with its corresponding secret key from the
credential repository. The precondition states that the $cert?$
must exists in the $certificates$ set.

\begin{schema}{RemoveCertificate0}
    \Delta CredentialRepository \\
    cert? : Certificate \\
 \where
    cert? \in  certificates \land  certificates' = certificates \setminus  \{ cert? \} \land \\
    key\_association' = \{ cert?.serial \} \ndres key\_association  \land \\
    cert\_association' = cert\_association \nrres \{ cert?.serial \}
\end{schema}

\section{Conclusion}
\noindent In this paper, we have presented a formal model of a VO architecture that combines PKI and username-password mechanisms in order to provide a usable security solution for VO end-users. The model uses a combination of state-based and event-based approach. The consistency of Z model is checked with the ZTC tool. The formalism has clarified the purpose of the VO, the explicit assumptions about sites, users, CAs and third
parties. The observation of the states enables to easily know the current capabilities of the user,  site and the VO. For instance, it becomes clear from the model of the user that he/she will have to maintain only one identity to authenticate to the gateway. Also, it makes it possible to find a user responsible for performing a task on the grid. This allows local sites and the entity running the gateway to monitor who access resources for auditing and billing purposes. Most importantly, we have linked the CSP model with the $Z$ specification in a systematic way. We derive the CSP interface of the VO from the \emph{Z} state, operations, precondition on each operation and initial state. \\

\noindent \textbf{Acknowledgment}: This work is supported by EPSRC through the User-Friendly Security for Grid Environments project (EP/D051754/1), RealityGrid platform (EPSRC EP/C536452/1), EU FP6 ViroLab (EU FP6 027446).

\bibliography{ws}
\bibliographystyle{plain}
\end{document}